%% file: main.tex
\documentclass[sigconf]{acmart}
\AtBeginDocument{%
  }

\setcopyright{acmlicensed}
\copyrightyear{2018}
\acmYear{2018}
\acmDOI{XXXXXXX.XXXXXXX}
\acmConference[Conference acronym 'XX]{Make sure to enter the correct
  conference title from your rights confirmation email}{June 03--05,
  2018}{Woodstock, NY}
\acmISBN{978-1-4503-XXXX-X/2018/06}

\copyrightyear{2025}
\acmYear{2025}
\setcopyright{rightsretained}
\acmConference[RecSys '25]{Proceedings of the Nineteenth ACM Conference on
Recommender Systems}{September 22--26, 2025}{Prague, Czech Republic}
\acmBooktitle{Proceedings of the Nineteenth ACM Conference on Recommender
Systems (RecSys '25), September 22--26, 2025, Prague, Czech
Republic}\acmDOI{10.1145/3705328.3759342}
\acmISBN{979-8-4007-1364-4/2025/09}

\begin{document}

\title{APS Explorer: Navigating Algorithm Performance Spaces for Informed Dataset Selection}

\author{Tobias Vente}

\orcid{0009-0003-8881-2379}
\affiliation{%
  \department{Adrem Data Lab}
  \institution{University of Antwerp}
  \city{Antwerp}
  \country{Belgium}
}
\email{tobias.vente@uantwerpen.be}

\author{Michael Heep}
\orcid{TODO}
\affiliation{%
  \department{Intelligent Systems Group}
  \institution{University of Siegen}
  \city{Siegen}
  \country{Germany}}
\email{michael.heep@student.uni-siegen.de}

\author{Abdullah Abbas}
\affiliation{%
  \department{Intelligent Systems Group}
  \institution{University of Siegen}
  \city{Siegen}
  \country{Germany}}
\email{abdullah.abbas@student.uni-siegen.de}

\author{Theodor Sperle}
\affiliation{%
  \department{Intelligent Systems Group}
  \institution{University of Siegen}
  \city{Siegen}
  \country{Germany}}
\email{theodor.sperle@student.uni-siegen.de}

\author{Joeran Beel}
\affiliation{%
  \department{Intelligent Systems Group}
  \institution{University of Siegen}
  \city{Siegen}
  \country{Germany}}
\email{joeran.beel@uni-siegen.de}

\author{Bart Goethals}
\orcid{0009-0003-8881-2379}
\affiliation{%
  \department{Adrem Data Lab}
  \institution{University of Antwerp}
  \city{Antwerp}
  \country{Belgium}
}
\email{bart.goethals@uantwerpen.be}

\renewcommand{\shortauthors}{TODO}




\input{content/abstract}

\maketitle

\input{content/introduction}
\input{content/related_work}
\input{content/method}
\input{content/aps_explorer}

\bibliographystyle{ACM-Reference-Format}
\bibliography{references}

\end{document}

%% file: content/abstract.tex
\begin{abstract}
Dataset selection is crucial for offline recommender system experiments, as mismatched data (e.g., sparse interaction scenarios require datasets with low user-item density) can lead to unreliable results. 
Yet, 86\% of ACM RecSys 2024 papers provide no justification for their dataset choices, with most relying on just four datasets: Amazon (38\%), MovieLens (34\%), Yelp (15\%), and Gowalla (12\%). 
While Algorithm Performance Spaces (APS) were proposed to guide dataset selection, their adoption has been limited due to the absence of an intuitive, interactive tool for APS exploration. 
Therefore, we introduce the APS Explorer, a web-based visualization tool for interactive APS exploration, enabling data-driven dataset selection.
The APS Explorer provides three interactive features: 
(1) an interactive PCA plot showing dataset similarity via performance patterns,
(2) a dynamic meta-feature table for dataset comparisons, and
(3) a specialized visualization for pairwise algorithm performance.

\end{abstract}

%% file: content/introduction.tex
\section{Introduction}

One of the most important decisions in preparing offline recommender system experiments is dataset selection \cite{10.1145/2090116.2090122, beel2024informed}. 
Different datasets can lead to varying trends in performance metrics, training and inference time, or hinder the interpretability of results \cite{10.1145/2090116.2090122, beel2024informed}. 
For example, research on sparse interaction scenarios requires datasets with low user-item density, or investigating the reduction of bias in datasets requires datasets with empirically present bias to avoid misleading conclusions.  
Consequently, dataset selection must be rigorously aligned with the experiment’s objectives to ensure the experiment's validity.

Yet, the reasoning behind dataset selection in recommender systems experiments remains largely unclear \cite{Cremonesi_Jannach_2021}. 
We conducted a survey of all full papers at the ACM RecSys 2024, which reveals that only 14\% (8 of 58) provide a justification for the suitability of their dataset selection. 
The remaining 86\% (50 of 58) offer no explanation on why their selected datasets are appropriate for their experiments.
Among the remaining papers, 50\% (29 of 58) justify their dataset selection by stating that they are ``real-world'' datasets, 20\% (12 of 58) state that they focus on public availability, and 16\% (9 of 58) rely on established popularity, such as the MovieLens datasets.

Additionally, researchers mainly limit their experiments to only four dominant datasets: Amazon (38\%, 22 of 58), MovieLens (34\%, 20 of 58), Yelp (15\%, 9 of 58), and Gowalla (12\%, 7 of 58). 
This heavy reliance on a narrow set of datasets, without clear experimental justification, suggests that the field may be overfitting to these benchmarks rather than selecting data that best fit their research questions. 
Prior work supports this observation \cite{BauerSaidLandscape2024, chin2022datasets, sun2020we, Sun2023DaisyRec2, energypaper}, raising concerns about the generalizability and validity of the findings in recommender systems research experiments.

To facilitate dataset selection in recommender systems, \citet{beel2024informed} introduced Algorithm Performance Spaces (APS). 
The APS framework represents datasets in an n-dimensional space, where each dimension corresponds to the performance of a distinct recommendation algorithm. 
Therefore, spatial distance between datasets in the APS quantifies their diversity; greater separation indicates more distinct differences in how algorithms perform across them.

Although \citet{beel2024informed} demonstrates how APS improves dataset selection by identifying diverse dataset, its practical adoption remains limited due to the lack of an intuitive, interactive tool for the exploration of APS. 
While \citet{beel2024informed} provide raw data in form of .csv files in their GitHub repository, researchers must manually download the files, modify the code, and generate PCA plots to visualize the data. 
This hinders accessibility, efficiency, and makes it difficult to use the provided data for an informed dataset selection.

We present the APS Explorer\footnote{\url{https://datasets.recommender-systems.com/}\label{fn1}}, a web-based visualization tool that enables researchers to leverage APS for informed dataset selection. 
The APS Explorer evaluates 29 recommendation algorithms across 96 datasets using nDCG, Recall, and Hit Ratio to construct a comprehensive APS. 
The web tool provides three interactive features: (1) a PCA visualization showing dataset similarity via performance patterns, (2) a dynamic meta-feature table for dataset comparisons, and (3) a specialized view for pairwise algorithm performance across recommender systems datasets.



%% file: content/related_work.tex
\section{Related Work}

Recommender systems datasets have been extensively studied. For example 
synthetic dataset generation \cite{syntheticdatasets2023,synthetic1,synthetic2,Grennan2019}, 
dataset augmentation techniques \cite{recDataaug1,recDataaug2,Edenhofer2019}, 
or pruning strategies \cite{BeelPruning2019d}.

However, only two prior studies \cite{chin2022datasets, beel2024informed} align with our goal of facilitating informed dataset selection. 
(1) \citet{beel2024informed} extended the Algorithm Performance Space (APS) framework, introduced by \citet{tyrrell2020algorithm}, to quantify dataset diversity in recommender systems. In this framework, each dimension represents the performance of a distinct algorithm, and datasets are positioned based on their performance patterns. The Euclidean distance between datasets in this space measures their diversity. Greater distances indicate more divergent algorithm performance.
(2) \citet{chin2022datasets} took a complementary approach, advocating for dataset selection based on intrinsic characteristics like sparsity, size, and interaction distribution. They clustered 51 datasets into five groups using k-means and evaluated five algorithms 
on representative datasets from each cluster. Their experimental results revealed statistically significant performance variations (p < 0.05), with algorithms like RP3beta showing up to 45\% better performance on moderately sized, sparse datasets compared to denser ones.

While the influence of dataset characteristics on algorithm performance has been acknowledged since at least 2012 \cite{dataimpactonalgorithm2012}, the relationship remains nuanced. For instance, similar user/item distributions or sparsity levels do not consistently predict algorithm behavior \cite{beelrepro2016,worryanalysis,Cremonesi_Jannach_2021,Thomas2011}. This inconsistency is exemplified by \citeauthor{chin2022datasets}'s finding that Amazon datasets 
appeared in different clusters yet exhibited no clear intra- or inter-cluster performance trends.

While existing research has thoroughly investigated various aspects of recommender systems datasets, the development of intuitive and easily accessible visualizations to support recommender systems dataset selection remains an open challenge. 


%% file: content/method.tex
\section{Method}
We constructed and analyzed an actual Algorithm Performance Space (APS) by evaluating 29 recommendation algorithms from RecBole \cite{recbole} across 96 recommender-system datasets using default hyperparameters. The dataset collection comprised 75 explicit-feedback and 21 implicit-feedback datasets, with all explicit feedback converted to implicit interactions (by treating each rating as positive) for consistent top-n recommendation generation. All datasets underwent 5-core pruning prior to analysis.

The experiments were conducted within a 7,000 GPU-hour budget on our university's NVIDIA Tesla V100 cluster, employing 5-fold cross-validation with a strict 30-minute training limit per algorithm-fold-dataset combination. While this approach using default hyperparameters and constrained training time may not yield optimal performance for every algorithm, we argue it provides a valid foundation for initial APS dataset characterization.

%% file: content/aps_explorer.tex
\section{The APS Explorer}

\begin{figure}
    \centering
    \includegraphics[width=\linewidth, height=250pt]{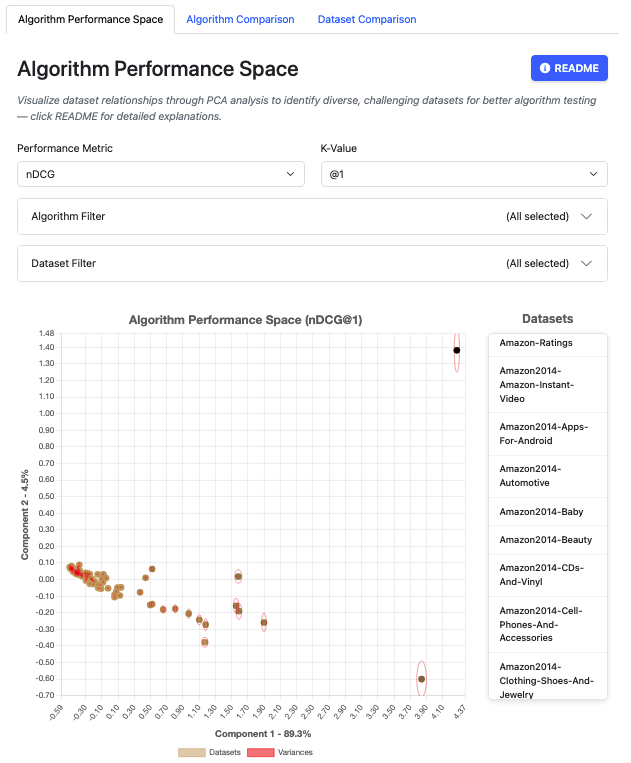}
    \caption{The APS Explorer UI}
    \Description[Test]{Test}
    \label{fig:stages}
    \vspace{-17pt}
\end{figure}

The APS Explorer\footref{fn1} enables researchers to systematically explore and select datasets, facilitating the identification of diverse datasets that align with their research objectives. This not only enhances the technical rigor of their work but also provides empirical grounding to strengthen dataset justification in publications and experiments. The APS Explorer is designed to be extensible, with ongoing development aimed at adapting to evolving researcher needs. Rather than imposing a new standard, the APS Explorer serves as a practical resource to improve dataset selection and transparency. The source code is publicly available on GitHub\footnote{\url{https://github.com/ISG-Siegen/APS-Explorer-Website}}.

The APS Explorer facilitates data-driven dataset selection through three core analytical modules, each accessible via dedicated tabs in the web-based tool: (1) the \textit{Algorithm Performance Space} for multidimensional exploration of algorithmic performance for a strategic dataset selection, (2) the \textit{Algorithm Comparison} tool for pairwise evaluation of recommender systems algorithm performance on recommender systems datasets, and (3) the \textit{Dataset Comparison} module for metadata-based benchmarking. All components support dynamic selection of evaluation metrics like nDCG, Hit Rate, Recall, and adjustable $K$-value parameters.

\textbf{\textit{Algorithm Performance Space:}}
The Algorithm Performance Space tab enables interactive exploration of the 29-dimensional performance space through two-dimensional PCA projections. 
By applying Principal Component Analysis (PCA) to high-dimensional performance vectors, which aggregate results from 29 algorithms across 96 datasets, the tool distills complex relationships into an interpretable 2D visualization. 
Researchers can interactively filter algorithms and datasets to compare their distributions in the reduced space and select a diverse subset suited for their analysis.

Furthermore, the first tab displays dataset difficulty, quantified using normalized positional quintiles. The difficulty score is computed as $\text{Difficulty Score} = (\text{norm}(C_1) + \text{norm}(C_2))/2$, with threshold quintiles ensuring an even distribution across five discrete difficulty levels. This allows researchers to not only assess dataset similarity but also evaluate how algorithms perform relative to different difficulty levels, providing insight into how dataset complexity influences algorithmic performance.

\textbf{\textit{Algorithm Comparison:}}
The Algorithm Comparison tab provides a visualization for pairwise algorithm evaluation. The plot partitions performance into quadrants by percentile thresholds: the lower-left quadrant (colored red) indicates mutual underperformance (bottom 25\% of results), while the upper-right quadrant (green) denotes joint strong performance (top 25\%). Asymmetric performance outcomes are shown in blue (Algorithm A superior) and yellow (Algorithm B superior) regions, with the central white area representing moderate performance (middle 50\%). This framework supports dynamic metric selection and $K$-value parameterization.

\textbf{\textit{Dataset Comparison:}}
The Dataset Comparison tab presents a tabular interface for systematic metadata analysis across all 96 datasets. Key comparative metrics include sparsity ratios, interaction distributions, and cold-start risk indicators. Researchers can filter, sort, and compare datasets to identify those best suited for their experiments. In addition, they can save selections or export metadata as .csv files for further analysis within the APS Explorer.